\pdfoutput=1

\documentclass{article}
\usepackage{spconf,amsmath,graphicx}
\usepackage{color}

\usepackage{algorithmicx}
\usepackage{algpseudocode}
\usepackage{algorithm}
\usepackage{caption}
\usepackage{multicol}
\usepackage{multirow}

\title{Calibration of fisheye camera using entrance pupil}
\name{Peter Fasogbon, Emre Aksu}
\address{Nokia Technologies, 33100 Tampere, Finland}
\begin{document}
%
\maketitle
\begin{abstract}
Most conventional camera calibration algorithms assume that the imaging device has a Single Viewpoint (SVP). This is not necessarily true for special imaging device such as fisheye lenses. As a consequence, the intrinsic camera calibration result is not always reliable.   In this paper, we propose a new formation model that tends to relax this assumption so that a Non-Single Viewpoint (NSVP) system is corrected to always maintain a SVP, by taking into account the variation of the Entrance Pupil (EP) using thin lens modeling. In addition, we present a calibration procedure for the image formation to estimate these EP parameters using non linear optimization procedure with bundle adjustment. From experiments, we are able to obtain slightly better re-projection error than traditional methods, and the camera parameters are better estimated. The proposed calibration procedure is simple and can easily be integrated to any other thin lens image formation model.	
\end{abstract}

\section{Introduction}
Camera calibration is the estimation of a camera's mapping function between a set of known world points and their measured image coordinates. The parameters that define this mapping are usually divided into two categories: intrinsic and extrinsic parameters.  The intrinsic parameters represent the internal characteristics of the image lens and sensors, while the extrinsic parameters model the relative position and orientation of the camera to the 3D world \cite{Zhang:2000}. 

\vspace{0.2cm}

\begin{figure}[htb]
	\begin{center}
		\includegraphics[height=5cm, width=6cm]{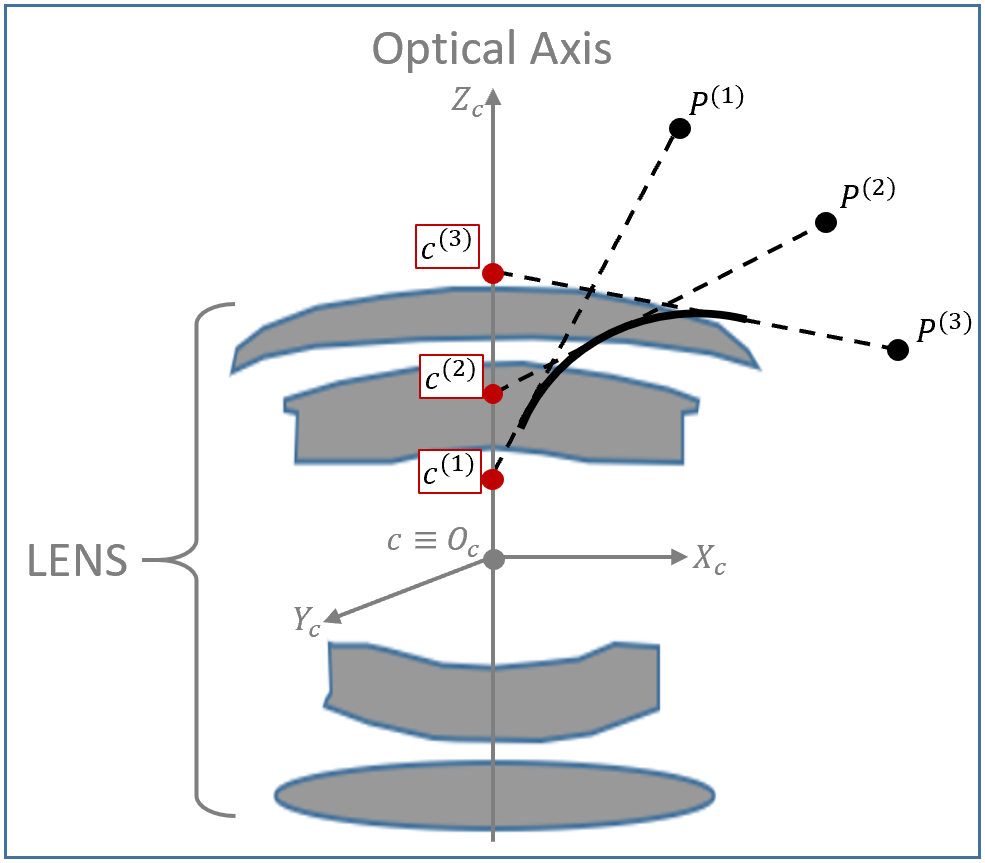}
	\end{center}

\vspace{-0.5cm}
	\caption{Illustration of the varying Entrance Pupil (EP). The entrance $c$ coincide with the optical axis $O_c$. The intersection of the incidence rays originated from world points $P^{(1)}, P^{(2)}, P^{(3)}$ intersect the optical axis at off-axis shift $c^{(1)}, c^{(2)}, c^{(3)}$ from the EP.
	}
	\label{fig:osznec}
\end{figure}

\begin{figure}[htb]
		\centering
		\centerline{		\includegraphics[height=7cm, width=6cm]{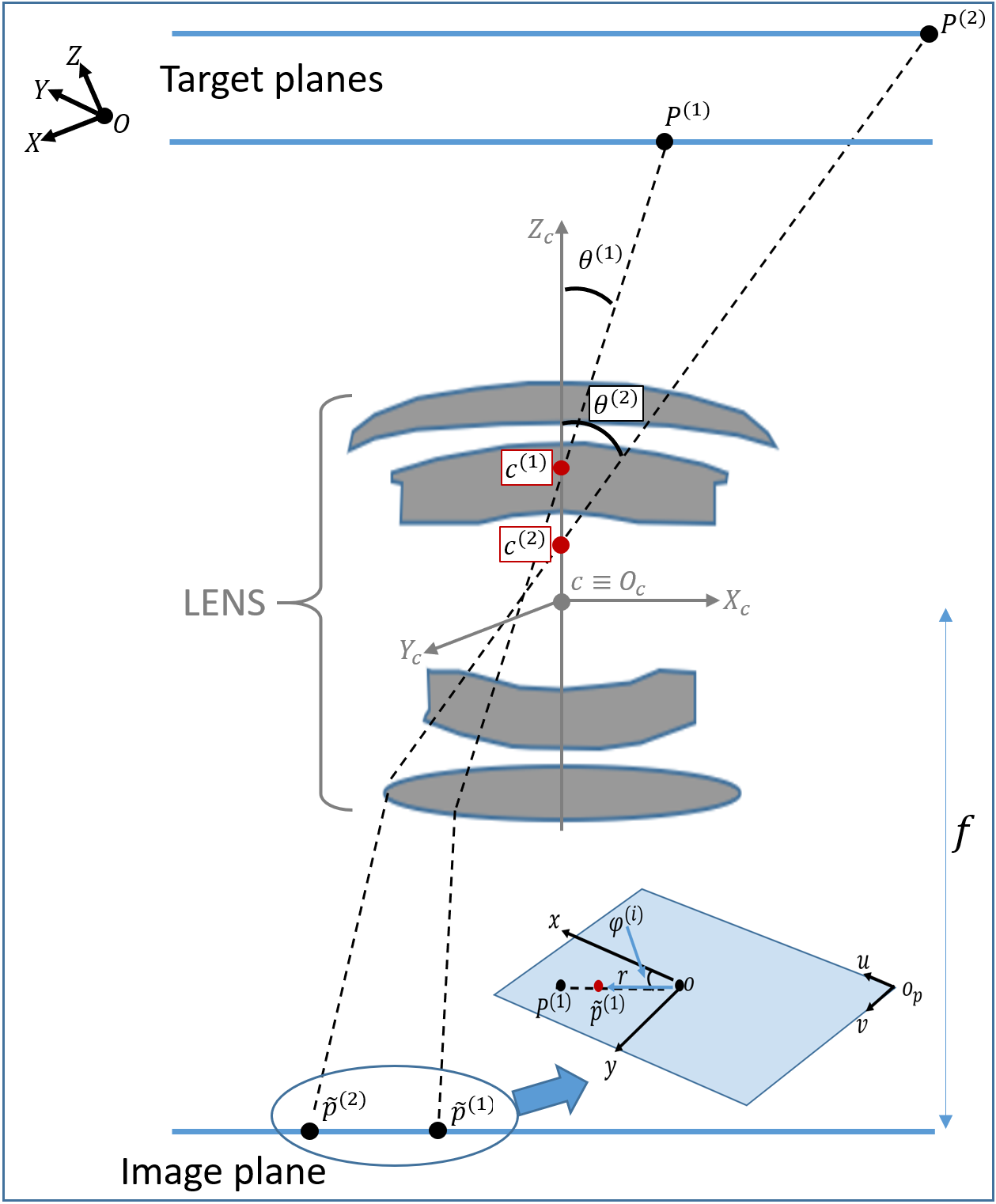} }
		

	\caption{ Image formation model for fisheye cameras under Non-Single Viewpoint (NSVP). $(P^{(1)}, P^{(2)})$ represent the world point projected onto the image plane at $(\tilde{p}^{(1)}, \tilde{p}^{(2)})$ in the presence of lens distortion. Assuming no lens distortion, the projected image points are $(p^{(1)}, p^{(2)})$. These point rays both form incidence angles $(\theta^{(1)}, \theta^{(2)})$ with the optical axis.  }
	\label{fig:fullpattfg}
\end{figure}

Most of the popular conventional imaging system assume that all incidence ray from a world point must pass through the optical axis in order to conform with a Single Viewpoint (SVP) cameras, under thin lens model \cite{Hall:1982}. However, the single viewpoint assumption is not always true since imaging system are made of various optical elements and/or mirrors. In figure \ref{fig:osznec}, we illustrate a Non-Single Viewpoint (NSVP) imaging using a fisheye lens. The locus of all the viewpoints in such system setup form a caustic depicted with thick curve line. The dotted lines represent the incidence ray originated from world points $(P^{(1)}, P^{(2)}, P^{(3)})$ with varying incidence angles from the optical axis. The incidence rays meet at off-axis shift $(c^{(1)}, c^{(2)}, c^{(3)})$ from the Entrance Pupil (EP) denoted as $c$ which is equivalent to the optical center $O_c$. Only the incidence angle that coincides with the  optical axis meets at the EP in the figure illustration. As the incidence angle increase away from the optical axis, the off-axis shift value increase as well.

For the majority of optical lenses with narrow field of view, the varying EP is small and thus can be ignored during the intrinsic camera calibration.  However, this is not the case for special imaging device such as catadioptric and fisheye lenses. A fisheye lens consists of several refractions and reflections in the imaging formation process as a result of large number of optical elements. Therefore, a single viewpoint assumption for fisheye camera calibration in \cite{Kannala06ageneric, fasogbon1:2018} implies an unrealistic constraint on the entrance pupil location, which impacts the parameter's sensitivity and correctness. There are literature that incorporate EP variation, such as \cite{Di:2004, Grossberg:2014, Swaminathan:2001, Kumar:2014} using thin lens and \cite{Kumar:2015} using thick lens for perspective cameras. Also, \cite{Gennery2006} using thin lens for fisheye cameras. Most of the thin lens model either have extreme complexity or more suitable for catadioptric imaging system only. 

\vspace{0.5cm}



A new camera model based on varying entrance pupil is proposed for fisheye lens calibration. We introduce EP correction strategy to bring NSVP to a SVP imaging model. In the proposed model, the extrinsic part of the model incorporates the EP parameters that are used to correct the incidence ray from world points, under a non-linear optimization through bundle adjustment procedure. The experimental validation on calibration data shows slightly better reprojection error over two representative state of the art \cite{Kannala06ageneric, Gennery2006} that are under SVP and NSVP categories respectively. 


\vspace{0.5cm}

In the next section, we present state of the art image formation model and necessary equations needed to understand the proposed method. The section starts with the equidistant projection formulation \cite{Kannala06ageneric}, followed by the optical lens modeling for fisheye lenses. The proposed image formation model and the suitable calibration procedure is detailed in section 3. In section 4, we perform experiments on calibration image data, and comparison with two representative state of art methods. In the last section, we summarize the paper and explain the future direction of this work.

\section{Image Formation Model}\label{sb:intrinsic}

In the figure \ref{fig:fullpattfg}, $i=1,\ldots, n$ represents the number of world points projected onto the lens.  $O(X,Y,Z), O_c(X_{c},Y_{c},Z_{c})$ represent the world and camera origins. $o(x,y), o_p(u,v)$ represent the image and pixel coordinates origin, $f$ is the camera's focal length and $r$ is the distance between a projected image point $p$ and origin $o$. In addition, $O_c$ correspond to the optical center and entrance pupil $c$ of the camera/lens, and $Z_{c}-$axis describe the camera's optical axis. In the figure, the incidence ray for every object point should pass through the optical center under SVP assumption for thin lens modeling.


%


\vspace{0.2cm}
\noindent
{\bf Extrinsic :} The camera point $P_c^{(i)}(X_c, Y_c, Z_c)$ is derived from world point $P^{(i)}(X, Y, Z)$ using equation (\ref{fig:p4}). The $3\times 3$ rotation matrix $\mathcal{R}$ and $3\times 1$ translational vector $\mathcal{T}$ are the extrinsic parameters that determine the external pose of a camera to the world coordinates \cite{Hall:1982, Zhang:2000}. Thanks to equation (\ref{fig:p4}), the incidence angle $\theta^{(i)}$ under SVP with the optical axis is defined as \resizebox{.35\hsize}{!}{$\theta^{(i)} = cos^{-1} \frac{Z_c^{(i)}}{\| \overrightarrow{ O_cP_{c}^{(i)}} \|}$}. 
\begin{equation}  
\begin{split}
\begin{bmatrix}
X_{c}\\
Y_{c}\\
Z_{c}
\end{bmatrix}^{(i)} = [\mathcal{R}|\mathcal{T}]  \begin{bmatrix}
X\\
Y\\
Z\\
1
\end{bmatrix}^{(i)}
\end{split}
\label{fig:p4} 
\end{equation}

\vspace{0.2cm}
\noindent
{\bf Intrinsic $\&$ Lens distortion :} 
In the presence of optical distortions and using equidistant projection \cite{Gennery2006, Kannala06ageneric, Hall:1982}, distorted image point $\tilde{p}^{(i)}(\tilde{x},\tilde{y})$ is expressed in equation (\ref{kansimp}), where angle $\varphi = tan^{-1} \frac{y}{x}$, radial distance $r(\theta) = \theta + k_1 \theta^3 + k_2 \theta^5  + k_3 \theta^7  + k_4 \theta^9$  \cite{Kannala06ageneric}, and $k_1, k_2, k_3, k_4$ are the lens distortion parameters.

\begin{equation}  
\begin{split}
\tilde{x} = r(\theta). cos (\varphi),\\
\tilde{y} = r(\theta). sin (\varphi) \\ 
\end{split}
\label{kansimp}
\end{equation}

As a final step, image point $\tilde{p}^{(i)}(\tilde{x},\tilde{y})$ is transformed to its pixel coordinate location as : $\tilde{u} = f_x.\tilde{x} + sk.\tilde{y} + u_0$  and $\tilde{v} = f_y.\tilde{x} + v_0$ respectively. The parameter $sk$ is the skew factor, $(f_x, f_y)$ are the focal lengths in pixels, $(u_0$, $v_0)$ are the principal point in pixels \cite{Zhang:2000}.

%

%

\section{Proposed Model using Entrance Pupil}
Some fisheye calibration methods \cite{Kannala06ageneric, Scaramuzza4059340, fasogbon1:2018} put a lot of unnecessary pressure on the optical lens model to simplify the distortions introduced as a result of the varying EP. This tend to affect the accuracy of the calibration and proper estimation of the distortion parameters. Most especially, closer object to the camera tends to be the most impacted by the entrance pupil variation. This problem is investigated using a 3D reconstruction of some measured object in an industrial application. As a result, we are motivated in this work to separate the EP formation from the intrinsic modeling of the system. With this approach, EP is then modeled as part of the camera extrinsic parameters.




\vspace{0.2cm}
\noindent
{\bf Extrinsic :} 
For the non-single viewpoint in Fig. \ref{fig:fullpattfg}, the incidence angle is calculated as $\theta^{(i)} = cos^{-1} \frac{Z_{c}^{(i)}}{\| \overrightarrow{c^{(i)}P_{c}^{(i)}} \|}$. The EP variation is then modeled as a function of $\theta$ by using higher degree of odd polynomial. This is a reasonable assumption when one assume there is no decentering in the lens system  \cite{Gennery2006, Brown71close-rangecamera}. Our approach ensure that the EP model is separated from image coordinate system during future optimizations, and thus is modeled as part of the extrinsic parameters. As a result, we do not need to worry about decentering effect.

\begin{figure}
	\begin{center}
		\includegraphics[height=5cm, width=6cm]{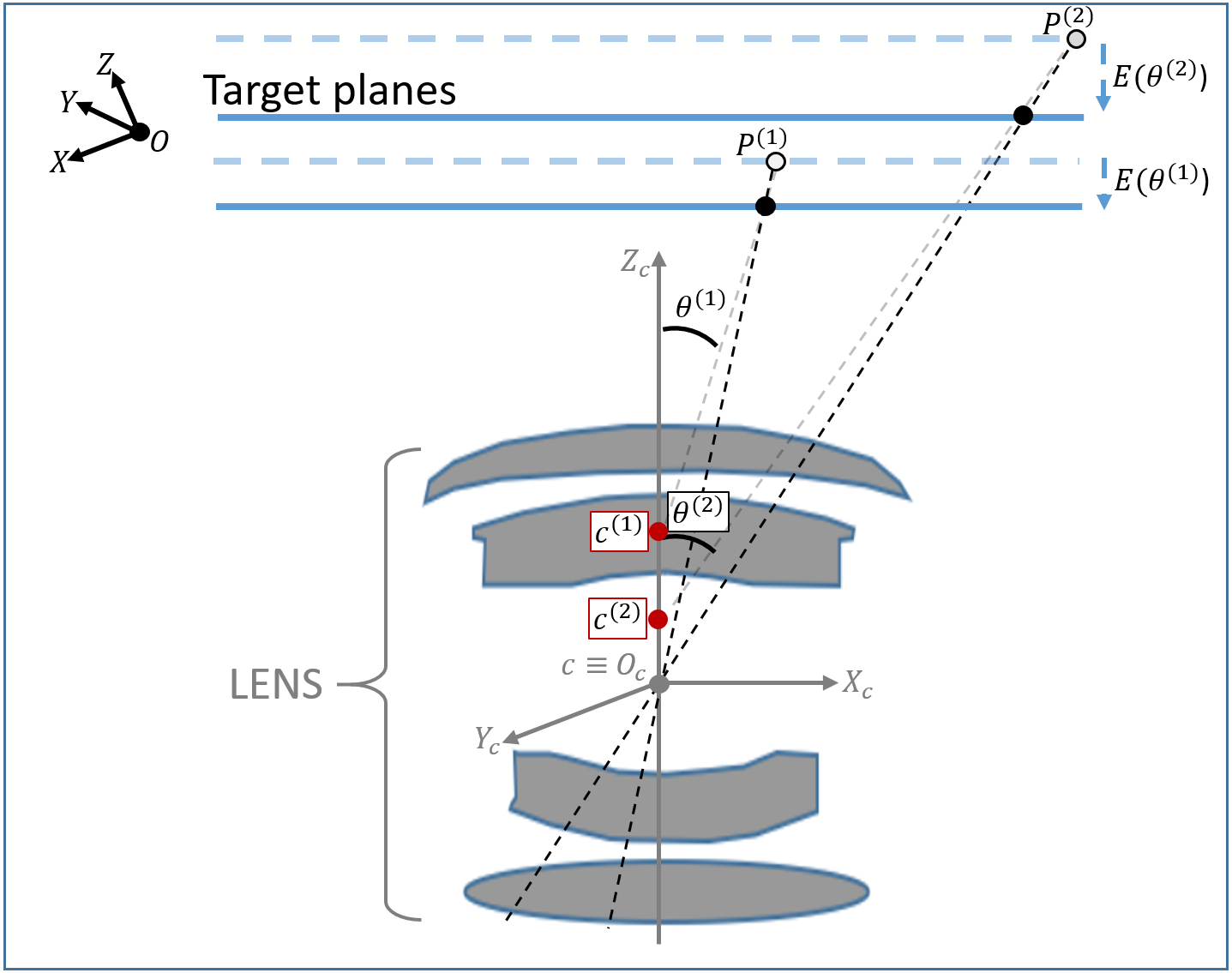}
	\end{center}
	\caption{Our correction for the entrance pupil variation. The world points $(P^{(1)}, P^{(2)})$ are moved in a series of optimizations to an optimal target plane using parameters  $E(\theta^{(1)})$ and $E(\theta^{(2)})$. 
	}
	\label{fig:onecsols}
\end{figure}

\vspace{0.2cm}
The proposed solution is better illustrated in Figure \ref{fig:onecsols}. We move the off-axis $c^{(i)}$ to optical center $O_c$ in a series of non-linear optimization. This is equivalent to moving point $P^{(i)}$ along the $Z_c-$axis using parameter $E(\theta^{(i)})$ which is $E(\theta)=e_1*\theta^3    +   e_2*\theta^5     +    e_3*\theta^7  +    e_4*\theta^9 $, where $e_1$, $e_2$, $e_3$, $e_4$ are the entrance pupil parameters. Using the parameter $E(\theta)$, our extrinsic model is rewritten as in equation (\ref{err:rep4}).

\begin{equation}  
\begin{split}
\begin{bmatrix}
X_c \\
Y_c \\
Z_c\\
\end{bmatrix}^{(i)} =
\begin{bmatrix}
\mathcal{R} & \mathcal{T} \end{bmatrix}\begin{bmatrix}
X \\
Y \\
Z + E(\theta)\\
1 \\
\end{bmatrix}^{(i)}
\end{split}
\label{err:rep4}
\end{equation}

\vspace{0.2cm}

These complete the proposed extrinsic model to correct the varying EP. The entrance pupil variation most impact the Z-axis, so the proposed correction is significant, while the X and Y-axis corrections are left to be corrected as part of radial distortion in section \ref{sb:intrinsic}. For lenses with large distortions other than radial ones, various models that incorporates tilt, tangential distortions can be easily integrated to this model \cite{Brown71close-rangecamera, fasogbon1:2018}

\subsection{Calibration Procedure}
For the calibration procedure, we follow the popular planar based calibration method in \cite{Zhang:2000}. A planar calibration target with known feature points that is moved at different poses from the camera to be calibrated is used. Let us denote $i=1,\ldots, n$ number of feature points, and $j=1,\ldots, m$ number of poses. A world point and its corresponding distorted image pixel can then be represented as $P^{(i,j)}$ and  $\tilde{p}^{(i,j)}$ respectively.

The calibration procedure consist of three main steps: (i) feature extraction \cite{fasogbon2:2018}, (ii) initial camera parameters estimation \cite{Kannala06ageneric},	(iii) Non-linear refinement of initial camera parameters with the proposed model. 

\vspace{0.2cm}
\noindent
{\bf Feature Extraction :} We exert the correspondence between points $P^{(i,j)}$ on the calibration target and the projected image  pixels $\tilde{p}^{(i,j)}(u, v)$ using the method proposed in \cite{fasogbon2:2018}. 

\vspace{0.2cm}
\noindent
{\bf Initial camera parameters estimation :} We retrieve some initial camera parameters: intrinsic $(f_x, f_y, u_0, v_0, k_1, k_2)$ and extrinsic $(\mathcal{R}_{(j)} , \mathcal{T}_{(j)})$ for each pose $j$  using method proposed in traditional fisheye calibration \cite{Kannala06ageneric}.

\vspace{0.2cm}
\noindent
{\bf Non-linear optimization with EP:} The remaining parameters that have not been initialized from earlier step are set to zero i.e ($sk=0$, $k_3, k_4=0$, $e_1, \ldots, e_4=0$). Finally, all these parameters are refined in equation (\ref{err:rep}) by minimizing the reprojection error between the true image pixels $\tilde{p}^{(i,j)}$ and the reprojected one $\hat{p}^{(i,j)}$.  

\begin{equation}  
\begin{split}
min \sum\limits_{j=1}^m\sum\limits_{i=1}^n\|\tilde{p}^{(i,j)} - \hat{p}^{(i,j)}(\hat{\mathcal{K}},\hat{\mathcal{R}}_{(j)},\hat{\mathcal{T}}_{(j)},\hat{e}_1,\ldots,\hat{e}_4, \\ \hat{k}_1, \ldots, \hat{k}_4)\|^2\\
\end{split}
\label{err:rep}  
\end{equation}
where $\hat{\mathcal{K}}:\{\hat{f}_x, \hat{f}_y, \hat{sk}, \hat{c}_x, \hat{c}_y\}$ are the estimated intrinsic parameters, $\{ \hat{k}_1,\ldots,\hat{k}_4 \}$ are the estimated radial distortion parameters, and $\{ \hat{\mathcal{R}}_{(j)},\hat{\mathcal{T}}_{(j)}, \hat{e}_1, \ldots, \hat{e}_4 \}$ are the estimated extrinsic parameters. 
The optimization is done with bundle adjustment using Levenberg-Marquardt's method \cite{Triggs:1999:BAM:646271.685629}. Please note that the world point is corrected at each iteration to give the estimated $\hat{p}^{(i,j)}$.

\section{Experiments}	
\begin{figure}
	\begin{center}
		\includegraphics[height=4cm, width=6cm]{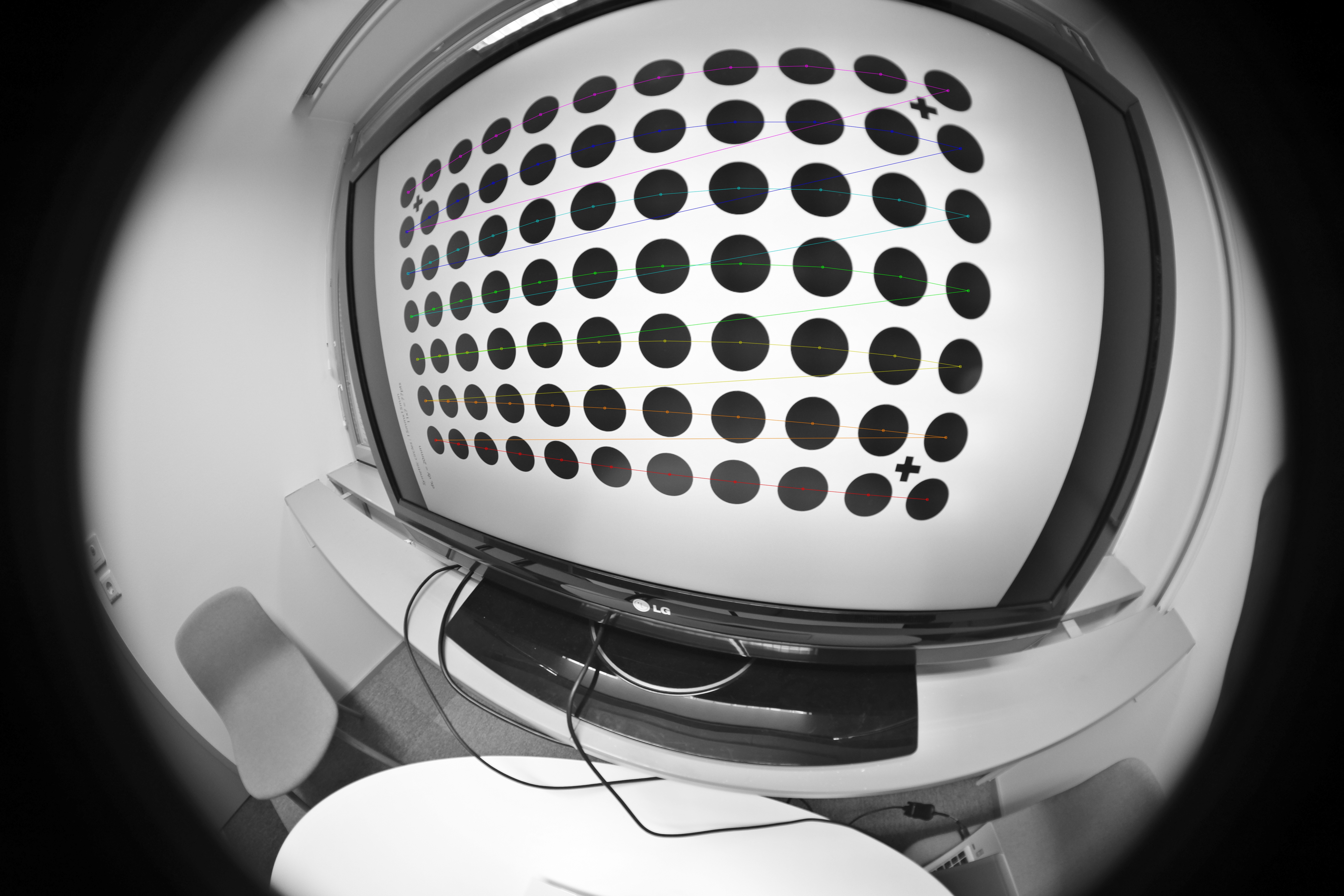}
	\end{center}
	\caption{An example image for pose $j$ of the calibration target. The extracted feature pixels are the center of gravity of the projected circular pattern.
	}
	\label{fig:oneds}
\end{figure}

The proposed calibration method has been evaluated on both real and synthetic dataset. The experiments have been done with a calibration target in a working distance of between $100 ~to~ 150~mm$ from the camera to be calibrated. This working distance is required to be able to accurately measure the capability of the system in an extreme setup. The calibration target is made up of (11$\times$7) 77 features in total (see figure \ref{fig:oneds}). The acquired image exhibit large optical distortion as a result of the fisheye lens. The feature pixels are the Center of Gravity (CoG) of the projected circular patterns, and have been extracted using the method proposed in \cite{fasogbon2:2018}.

The name of the fisheye lens camera device used for the calibration test is witheld due to intellectual property issue. However, it is a virtual reality device that is made up of 8 cameras with spatial resolution of 2048 $\times$ 2048. Each fisheye lens has a Field of View (FOV) of 195 $^{\circ}$ and focal length of 3.5mm. We have made several test on all of the 8 cameras but only the result of one principal camera is documented in this paper due to page limitation. The proposed calibration is evaluated with Root Mean Square (RMS) error between the true image pixels and reprojected ones using the estimated camera parameters, over 25 calibration images.  

\vspace{0.2cm}
In the remaining part of this section, we present the results of the proposed calibration method and compare the re-projection error with the representative state of art algorithms: (i) Kannala et al.  \cite{Kannala06ageneric} that is based on SVP approach and (ii) Gennery \cite{Gennery2006} that is based on NSVP approach. We have chosen Kannala et al.  \cite{Kannala06ageneric} as the compared fisheye calibration method over Scaramuzza et al. \cite{Scaramuzza4059340} because it provides  better result under our current experimental setup \cite{fasogbon2:2018}. All the evaluated methods are based on thin lens model which is the main focus in this work.

\vspace{0.2cm}
\noindent
{\bf Analysis of real image test:} The result and comparison between the experimented state of art methods and our approach are shown in Table \ref{tab:dfgggadf}. Different imaging conditions have been abbreviated in the caption of the table. The estimate of the principal point for our method is not very far from the physical principal point ($\frac{(u_0, v_0)}{2}$), even with the incorporation of the entrance pupil variation. 

By analysing the whole table, one can see that our calibration method provides the lowest mean reprojection error. The reprojection error ({\color{blue} 0.1956 }) is only slightly lower than Gennery method \cite{Gennery2006} but still validates the proposed method.

\begin{table}[htb]
	\begin{center}
		
		\resizebox{0.85\hsize}{!}{$	\begin{tabular}{|   l | c    || c  |  |c||}
			\hline     
			& {\bf Kannala \cite{Kannala06ageneric}} &  {\bf Gennery \cite{Gennery2006}}  &  {\bf Our Method} \\ 	 							
			\hline     \hline

			$(f_x)$	&	587.9521   & 597.3014 & 591.7301 \\  
			$(f_y)$	&593.1520	&   594.0581  & 592.0340 \\  \hline
			
			$sk$	& 0.2314	&  0.1526  & 0.1978 \\ \hline
			
			$(u_0)$	& 1020.3841	 & 1013.7982  & 1013.2001 \\ 		
			$(v_0)$	& 1028.0564	&  1024.2808   & 1025.0300 \\ \hline		
			
			$e_1$	& -	& 0.0712  & 0.0851\\  
			$e_2$	& -	& -0.1092  & -0.2577\\  
			$e_3$	& -	& 0.2917  & 0.3016\\  
			$ e_4$	&  - & 0.4861   & 0.5368 \\  \hline
			
			$k_1$	& 0.0152	&  -0.0096 & 0.0109\\  
			$k_2$	& 0.0161	&  0.0028 & -0.0013\\  
			$k_3$	& -0.0232	&  -0.0132 & 0.0008\\  
			$ k_4$	&  0.0065 &   0.0027 & -0.0004 \\  \hline
			
			$error$		& 0.2913 &  0.2115 & 0.1956 \\
			$std$	& 0.0891	&  0.0792   & 0.0735\\ \hline				
			
		\end{tabular} $}
	
\end{center}

\caption{The experimental results and comparism between the proposed calibration method (ours) and state of art methods  under SVP and NSVP.  ($error$- Mean error, $std$ - standard deviation error in pixels). }
\label{tab:dfgggadf}
\end{table}





\vspace{0.5cm}
\noindent
{\bf Analysis of synthetic test:} To demonstrate further the validity of the proposed model, we made synthetic experiment validation using exact setup under the real image test. We use synthetic generated feature points in order to eliminate any bias from circular feature detection under camera calibration. Table  \ref{tab:dfggxsdf} shows the standard deviation of the calibration parameters. In the table, we summarize the distortion parameters (radial and EP), so we only display their mean values. The smallest mean values between the compared methods for each camera parameter is illustrated in {\bf bold} font. From the outcome of the experiments, our method provides the lowest standard deviation of all the compared methods for the focal length and principal point estimation. 	

\begin{table}[htb]
	\begin{center}
		
		\resizebox{0.95\hsize}{!}{$	\begin{tabular}{|   l | c    || c  |  |c||}
			\hline     
			& {\bf Kannala \cite{Kannala06ageneric}} &  {\bf Gennery \cite{Gennery2006}}  &  {\bf Our Method} \\ 	 							
			\hline     \hline

			$(f_x, f_y)$	&	0.1145   &  0.9275 & {\bf 0.9081} \\    \hline
			
			$sk$	& {\bf  0.1982}	& 0.2140   & 0.2047 \\ \hline
			
			$(u_0, v_0)$	& 0.3023	 & 0.2312  & {\bf 0.2147} \\ 	 \hline		
			
			$e_1,\ldots, e_4$	& -	& 0.0313  & {\bf 0.0305}\\   \hline
			
			$k_1,\ldots, k_4$	& {\bf 0.0298}	& 0.0311  & 0.0354\\   \hline

		\end{tabular} $}
	
\end{center}

\caption{Mean standard deviation of the camera parameters estimated using synthetic data feature points. The smallest mean values is illustrated in {\bf bold} font.}
\label{tab:dfggxsdf}
\end{table}

\section{Conclusion}
We propose a new camera model that integrates entrance pupil variation for fisheye camera calibration. We model the entrance pupil variation as part of extrinsic and thus separate it from the intrinsic that includes the conventional lens distortion models. The proposed method gives accurate entrance pupil and other camera parameters estimate when applied on both real and synthetic images. 

By taking into account the Entrance Pupil variation, one ensure that the pixels located at the highly distorted extremities of the fisheye lens, and especially image objects that are very close to the lens are corrected. This improves the calibration result and it is expected to greatly improve future 3D reconstruction.


\bibliographystyle{IEEEbib}
\bibliography{strings,refs}

\end{document}